\documentclass[fleqn,10pt]{wlscirep}
\usepackage[utf8]{inputenc}
\usepackage[T1]{fontenc}
\usepackage{dcolumn}
\usepackage{textcomp}
\usepackage{lineno}

\usepackage{listings}
\usepackage{xcolor}
\title{AI Urban Scientist: Multi-Agent Collaborative Automation for Urban Research}

\author[1]{Tong Xia}
\author[2]{Jiankun Zhang}
\author[2]{Ruiwen You}
\author[2]{Ao Xu}
\author[2]{Linghao Zhang}
\author[2]{Tengyao Tu}
\author[2]{Jingzhi Wang}
\author[3]{Jinghua Piao}
\author[3]{Yunke Zhang}
\author[3]{Fengli Xu}
\author[3]{Yong Li}
\affil[1]{Vanke School of Public Health, Tsinghua University, Beijing, China}
\affil[2]{Zhongguancun Academy, Beijing, China}
\affil[3]{Department of Electronic Engineering, Tsinghua University, Beijing, China}

\begin{abstract}
Urban research aims to understand how cities operate and evolve as complex adaptive systems. With the rapid growth of urban data and analytical methodologies, the central challenge of the field has shifted from data availability to the integration of heterogeneous data into coherent, verifiable urban knowledge through multidisciplinary approaches.
Recent advances in AI, particularly the emergence of large language models (LLMs), have enabled the development of AI scientists capable of autonomous reasoning, hypothesis generation, and data-driven experimentation, demonstrating substantial potential for autonomous urban research. However, most general-purpose AI systems remain misaligned with the domain-specific knowledge, methodological conventions, and inferential standards required in urban science.
Here, we introduce the AI Urban Scientist, a knowledge-driven multi-agent framework designed to support autonomous urban research. Grounded in hypotheses, peer-review feedback, datasets, and research methodologies distilled from large-scale prior studies, the system constructs structured domain knowledge that guides LLM-based agents to automatically generate hypotheses, identify and integrate multi-source urban datasets, conduct empirical analyses and simulations, and iteratively refine analytical methods. Through this process, the framework synthesizes new insights in urban science and accelerates the urban research lifecycle.
\end{abstract}

\begin{document}

\flushbottom
\maketitle
\thispagestyle{empty}

\section*{Main}

More than half of the global population live in cities and therefore it is crucial to understand the dynamics of cities and diagnose emerging urban challenges, informing policy decisions on climate resilience, public health, economic development, and sustainability~\cite{acuto2018building,bettencourt2021introduction}. Yet despite rapid growth in data availability from remote sensing to mobility traces, and steady advances in statistical and computational methods, urban research workflow remains highly manual and fragmented. Researchers typically begin with observation-driven ideas, search for relevant datasets using ad hoc heuristics, manually assemble heterogeneous data sources, write extensive custom code across multiple languages, and iterate on results with limited reproducibility. These challenges inhibit systematic discovery, constrain hypothesis generation, and slow the translation of urban data into actionable insights.

Recent progress in large language models (LLMs) and autonomous agents has led to the emergence of the \textit{AI Scientist} paradigm~\cite{boiko2023autonomous,lu2024ai_scientist}: An AI scientist refers to an artificial intelligence–based system designed to conduct scientific research by executing the core steps of the scientific method with minimal human intervention. Its objective is to autonomously perform the iterative cycle of hypothesis formulation, solution or experimental design, experiment execution, result evaluation, and adaptive refinement, and in some cases to generate complete scientific manuscripts. Through feedback-driven learning, an AI scientist continuously improves its reasoning and experimental strategies, analogous to how human scientists iteratively test and refine hypotheses. In this sense, AI scientists represent a conceptual shift in artificial intelligence from computational tools that assist scientific tasks toward systems capable of autonomous scientific discovery.

Such systems also hold the potential to accelerate urban research by integrating search, reasoning, coding, and related capabilities. However, general-purpose AI Scientist frameworks—largely developed for fields such as machine learning, chemistry, or physics—are not directly applicable to urban studies. Urban phenomena are context-dependent, data are multi-source and heterogeneous, and analytical workflows require interdisciplinary knowledge spanning statistics, geospatial modeling, economics, epidemiology, and environmental science. Existing AI reviewer agents trained on ML corpora may misjudge high-quality urban study papers as “not novel enough,” reflecting a fundamental mismatch in epistemic standards. Similarly, existing data agents cannot extract datasets that are mainly introduced by scientific articles, which represent some of the most important resources in urban research.

\begin{figure*}[t]
    \centering
    \includegraphics[width=0.9\linewidth]{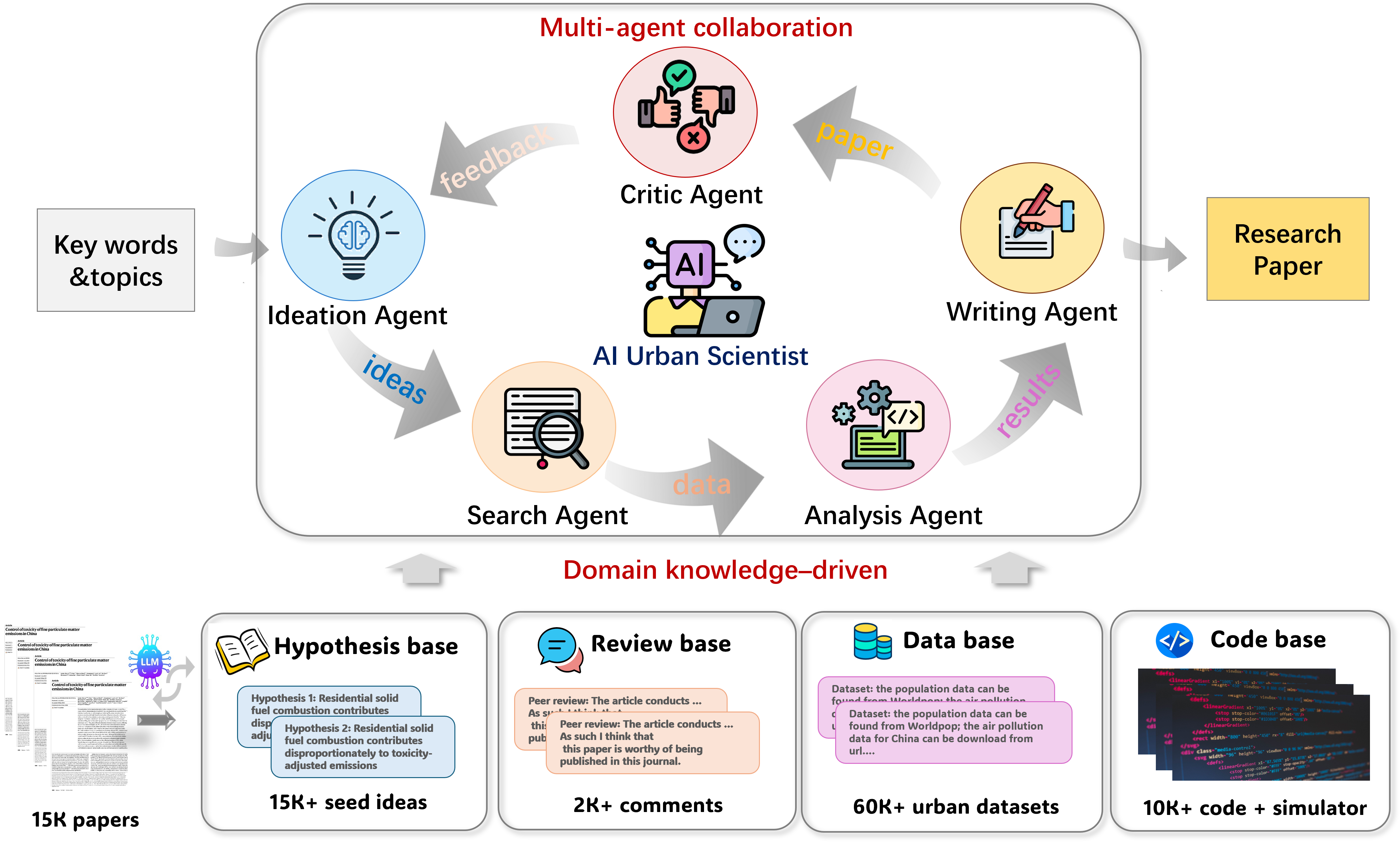}
  \caption{
\textbf{Overview of the AI Urban Scientist system.}
The system supports autonomous urban research by automating each stage of the scientific workflow, including identifying key topics, generating research hypotheses, discovering and integrating relevant datasets, conducting empirical analyses and simulations, and drafting the final research manuscript.
To enable this process, the system is grounded in four domain-specific knowledge bases:
(1) a hypothesis base distilled from over 15,000 academic papers,
(2) a review base comprising more than 2,000 expert peer-review comments,
(3) a data base containing over 60,000 urban datasets, and
(4) a code base of more than 10,000 analytical scripts and simulators reflecting standard urban modeling practices.
These knowledge resources coordinate five collaborative agents:
an \textit{Ideation Agent} for hypothesis generation and mutation,
a \textit{Search Agent} for multi-source data discovery,
an \textit{Analysis Agent} for executing analytical workflows,
a \textit{Writing Agent} for synthesizing results into a research paper,
and a \textit{Critic Agent}, fine-tuned on review knowledge, for evaluating research quality and providing iterative feedback.
}
    \label{fig:1}
\end{figure*}

To address these gaps, we propose AI Urban Scientist (Figure~\ref{fig:1}), a domain-informed, multi-agent system designed specifically for urban research.  
At its core, the AI Urban Scientist is designed as a knowledge-driven, multi-agent system in which specialized agents collaborate to emulate the full scientific cycle in urban research. Instead of relying on a monolithic LLM, the system decomposes the research process into interacting agents responsible for hypothesis generation, data discovery, empirical analysis, result synthesis, and critical evaluation. These agents are coordinated through structured domain knowledge distilled from prior literature, peer-review practices, urban datasets, and standard analytical workflows, enabling iterative feedback and role-specific reasoning. By explicitly modeling scientific research as a collaborative and self-correcting multi-agent process aligned with established urban study norms, the AI Urban Scientist supports autonomous yet verifiable urban science and advances scalable and insightful urban research.

Beyond the autonomous workflow, we develop an open platform (Figure~\ref{fig:8}) that allows researchers to directly use individual agents, experiment with full pipelines, and contribute new tools, datasets, and analytical components. This community-driven ecosystem aims to establish shared standards for hypotheses, datasets, and analytical practices in urban science—accelerating research and improving reproducibility.

In summary, our contributions are threefold:
\begin{itemize}

\item We present the first \emph{AI Urban Scientist} system that enables the entire urban research lifecycle—from problem formulation and hypothesis generation to data discovery, empirical analysis, and manuscript drafting—through autonomous and collaborative multi-agent coordination.

\item We propose a knowledge-driven multi-agent architecture grounded in structured urban science knowledge bases, including prior literature, peer-review practices, datasets, and analytical workflows, which systematically enhances domain alignment, methodological rigor, and research reliability.

\item We develop an open and extensible research platform that supports community use and further extension, enabling urban researchers to explore scientific questions in a more systematic, scalable, and collaborative manner.

\end{itemize}

\section*{Opportunities brought by AI Scientist}



\subsection*{What Is an AI Scientist and Why Does It Matter?}

An AI Scientist refers to an AI system designed to autonomously participate in the scientific discovery process by performing core research functions that traditionally rely on human expertise. Unlike conventional AI tools that support isolated tasks such as prediction or data analysis, an AI Scientist operates across the full scientific workflow, including literature synthesis, hypothesis generation, experimental design, data-driven reasoning, and iterative refinement of scientific knowledge.

The defining advantage of an AI Scientist lies in its ability to overcome fundamental human bottlenecks in scale, cognition, and coordination:

First, AI Scientist systems can systematically process and synthesize vast bodies of scientific literature at a speed and scope far beyond human capacity. By scanning tens of thousands of publications across multiple disciplines, these systems can identify latent patterns, unresolved contradictions, and underexplored research directions, enabling the generation of large numbers of testable hypotheses—including non-intuitive hypotheses that may be overlooked due to human cognitive biases or disciplinary conventions.

Second, AI Scientist systems excel at navigating high-dimensional and complex research spaces. In scientific settings characterized by large parameter spaces and interacting variables, human intuition often proves inefficient or unreliable. Through methods such as Bayesian optimization, reinforcement learning, and large-scale simulation, AI Scientist systems can autonomously design, optimize, and adapt experimental strategies, efficiently exploring vast search spaces that would otherwise be intractable.

Third, AI Scientist systems are inherently suited for large-scale pattern recognition and model construction in complex data environments. Modern scientific datasets are increasingly high-dimensional, noisy, and heterogeneous, posing significant challenges to human reasoning. AI systems can uncover nonlinear relationships and deep structural regularities across diverse data modalities, automatically constructing predictive and explanatory models that extend beyond human perceptual limits.

A further distinguishing feature of AI Scientist systems is their capacity for massively parallel and collaborative intelligence. Rather than functioning as a single monolithic agent, an AI Scientist can instantiate large populations of interacting agents with diverse reasoning strategies, enabling parallel exploration of hypotheses, analytical approaches, and interpretations. This form of scalable collective intelligence far exceeds the coordination capacity of human research teams.

In addition, AI Scientist systems are not constrained by social, institutional, or reputational pressures that shape human scientific behavior. Free from academic hierarchies, prevailing paradigms, and funding incentives, these systems can explore unconventional or counterintuitive research paths without psychological or professional cost, potentially opening avenues for disruptive discovery.

Finally, AI Scientist systems possess a unique capacity for continuous learning and deep interdisciplinary integration. Their knowledge bases can be updated in real time as new scientific results emerge, enabling cumulative and non-depreciating learning at a global scale. Moreover, by integrating knowledge across traditionally siloed disciplines, AI Scientist systems can reason from genuinely interdisciplinary perspectives, supporting scientific inquiry that transcends individual domain expertise.

Taken together, these capabilities define the AI Scientist not merely as a faster analytical tool, but as a new class of scientific agent—one that augments human intelligence by operating at scales, depths, and levels of integration that are fundamentally unattainable by human researchers alone.

\subsection*{How Can AI Scientist Help with Urban Research}


The AI Scientist framework offers transformative opportunities for urban research by reshaping each step of the scientific workflow—from problem discovery to hypothesis building, data acquisition, analysis, and publication. As depicted in Figure~\ref{fig:1}, AI agents can operate as research collaborators capable of synthesizing knowledge, automating analysis, and bridging disciplinary divides \cite{Chan2023LLMScience,AI2024NatureSci}. Below, we outline several key ways in which AI Scientist systems can advance urban research.

\begin{itemize}

 \item \textit{Accelerating Ideation}. Problem Discovery and Hypothesis Formation.
urban research has long suffered from a narrow band of canonical research problems and a heavy reliance on expert intuition. AI Scientist agents can explore large corpora of literature, extract underexplored topics, generate novel research directions, and synthesize interdisciplinary theories \cite{Liu2024DeepResearch}. These systems can identify emerging urban challenges—ranging from environmental risks to socio-spatial inequalities—and propose testable hypotheses at a pace and breadth far beyond human capability. This capability is particularly valuable in urban settings, where questions often span geography, sociology, economics, environmental science, and engineering.

 \item \textit{Enhancing Interdisciplinary Knowledge Integration.}
Urban phenomena are inherently cross-disciplinary, yet integrating theories from disparate fields is one of the most persistent bottlenecks in urban research. LLM-based AI Scientist systems excel at retrieving, summarizing, and reconciling knowledge across the social sciences, environmental sciences, public health, and computational modeling \cite{Park2023SocialAISystems}. By contextualizing a hypothesis within multiple disciplines, AI Scientist supports more comprehensive reasoning and fosters intellectually diverse urban inquiry.

 \item \textit{Comprehensive and Automated Data Search.}
Identifying suitable datasets is often one of the most time-consuming steps in urban research. AI agents can perform broad data discovery across open data portals, scientific archives, satellite sources, and sensor datasets. Recent developments in LLM-guided dataset retrieval \cite{Narayan2024AIAgentDS} allow the AI Scientist to match hypotheses with relevant data sources, assess data fitness, and even suggest complementary datasets to enrich analysis. This greatly reduces the manual burden and expands the scope of empirical studies.

 \item \textit{Automated Experimental Design and Data Analysis.}
Through auto-coding and LLM-to-code translation capabilities \cite{Jain2024AutoCoding}, AI Scientist systems can automatically generate analytical pipelines—ranging from causal inference models and spatial econometrics to simulation workflows and machine learning pipelines for environmental prediction. This not only accelerates research execution but also increases reproducibility by producing self-documented code, structured outputs, and standardized evaluation procedures. In urban research—where data can be noisy, high-dimensional, or incomplete—AI Scientist systems can also propose alternative analyses or robustness checks autonomously.

\item \textit{End-to-end research workflow with closed-loop improvement.}
Beyond autonomously generating research ideas and conducting data-driven analyses, the AI Scientist can produce structured scientific feedback that explicitly identifies weaknesses in research design and analytical strategies. This feedback is then leveraged to iteratively refine hypotheses, experimental setups, and analytical pipelines, forming a closed-loop process of continuous improvement across the entire research workflow.

 \item \textit{Providing Reusable Tools and Accelerating Community Building.}
A major limitation of current urban research is the lack of shared computational tools and standardized workflows. AI Scientist agents can help create modular, reusable codebases, automated templates for data processing, and shared analytical components that reduce redundancy across research groups. This mirrors the open-source tradition of machine learning but brings it into the urban research domain, fostering community-building and improving reproducibility across studies \cite{Khoo2024AutonomousScience}.

\end{itemize}

In summary, AI Scientist systems offer a powerful means to overcome the structural challenges of contemporary urban research. By accelerating ideation, enabling interdisciplinary synthesis, enhancing data accessibility, automating analyses, and fostering tool-sharing, AI Scientist has the potential to transition the field from a manual, expert-driven paradigm toward a more systematic, scalable, and discovery-oriented mode of scientific inquiry.

\section*{Towards Reliable AI Urban Scientist}



While recent progress in general-purpose AI scientist systems demonstrates impressive capabilities in hypothesis generation, literature synthesis, code writing, and autonomous experimentation, these systems remain insufficient for advancing the domain of urban research. The fundamental reason is that urban research differs from physics, machine learning, or chemistry in its theoretic development, data structures, methodological toolkit, and epistemic culture. Urban research requires extensive domain expertise to interpret heterogeneous datasets, understand contextual socioeconomic mechanisms, and apply appropriate analytical frameworks. Current AI Scientist agents lack these capabilities.

A striking example lies in state-of-the-art AI reviewers trained on machine learning peer-review corpora such as OpenReview. These agents can evaluate novelty and methodological rigor within ML domains, yet they fail when applied to interdisciplinary urban research. For instance, such models may incorrectly judge a Nature-level urban research paper as lacking novelty simply because it does not introduce a new algorithmic method—a misunderstanding rooted in their training on ML evaluation criteria, not on the epistemology of urban research. This mismatch illustrates that general-purpose AI reviewers and AI scientists are fundamentally misaligned with the intellectual standards and knowledge structures of urban research.

Therefore, building a reliable AI Urban Scientist requires more than scaling LLMs or improving autonomous reasoning. It demands the deliberate injection of domain knowledge, disciplinary conventions, and contextual understanding, much like how human experts develop deep intuition only after reading thousands of papers, internalizing canonical hypotheses, understanding available datasets, and mastering analytical tools.

This motivates the design of a knowledge-driven AI Urban Scientist, as illustrated in Figure~\ref{fig:2}. The system integrates a large hypothesis base distilled from 15K academic papers, a review base built from expert commentary, a curated data base of 20K+ urban datasets, and a code base and simulator representing standard analytical practices. Through coordinated agents—an Ideation Agent, Critic Agent, Data Agent, and Analyzing Agent—the system mirrors the workflow of a high-level human scientist: generating hypotheses, refining them through critique, identifying relevant datasets, executing analytical pipelines, and synthesizing insights. By embedding domain knowledge into each stage, the AI Urban Scientist is able to think with the depth, specificity, and contextual awareness required for credible urban research.

\subsection*{Automatic Urban Research Ideation}

\begin{figure*}[t]
    \centering
    \includegraphics[width=0.99\linewidth]{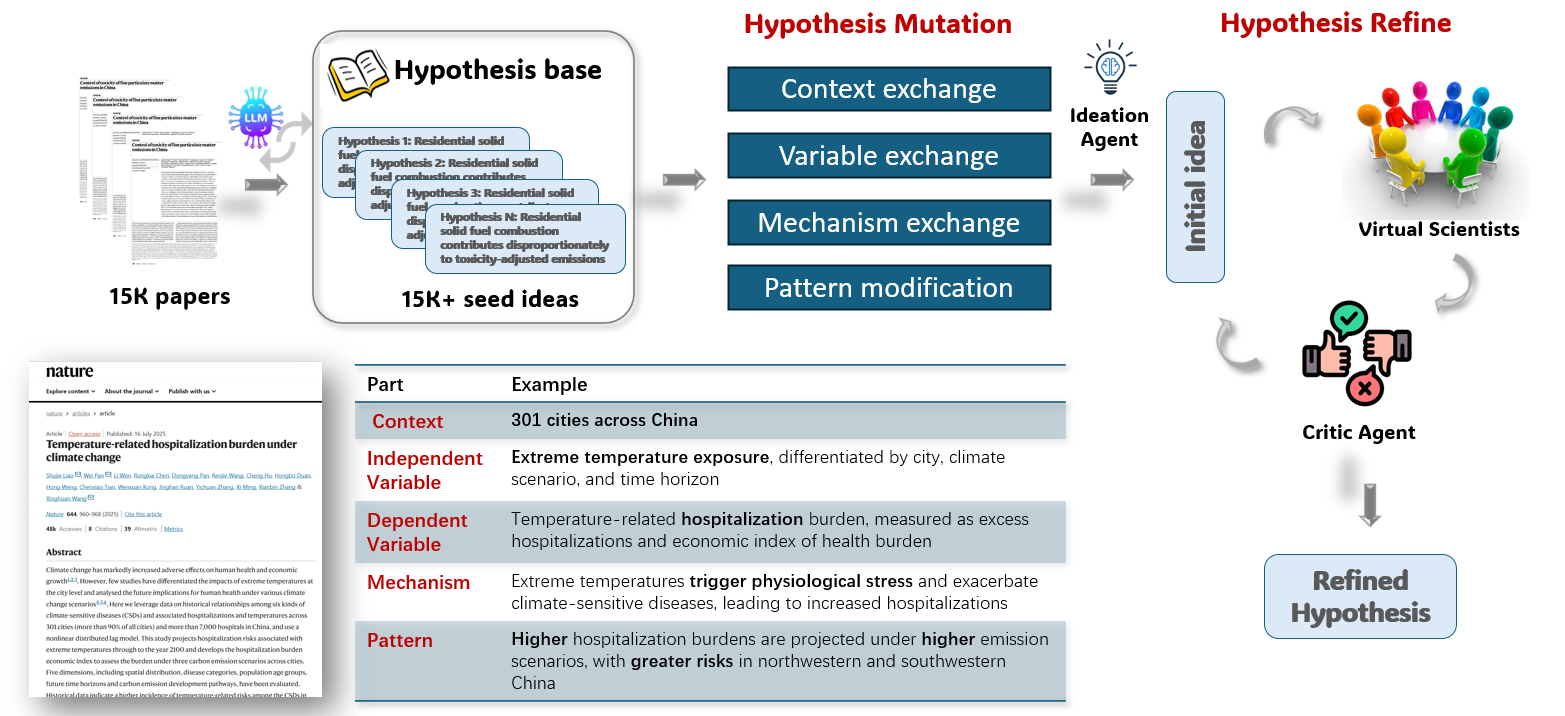}
    \caption{\textbf{The hypothesis-generation workflow of the Ideation Agent.}
The system constructs a hypothesis base from 15K research papers and decomposes mature hypotheses into CAMP components—Context, Variables, Mechanism, and Pattern. The Ideation Agent applies four scientific transformations (context exchange, variable exchange, mechanism exchange, and pattern modification) to generate initial hypotheses. These ideas undergo iterative refinement through two mechanisms: a panel of multi-disciplinary virtual scientists and a domain-trained Critic Agent. The resulting refined hypotheses form high-quality candidates for downstream empirical analysis. }
    \label{fig:3}
\end{figure*}

To generate high-quality and domain-grounded hypotheses for urban research, our system introduces the Ideation Agent, a knowledge-driven module designed around the CAMP theoretical framework. In urban research, a mature and scientifically testable hypothesis can typically be decomposed into five essential components:

\begin{itemize}

 \item \textbf{C (Context)}: the spatial, temporal, or socioeconomic setting in which the hypothesis is situated.

 \item \textbf{A \& B (Independent and Dependent Variables)}: measurable factors that establish the core relationship to be tested.

 \item \textbf{M (Mechanism)}: the causal pathway linking variables A and B.

 \item \textbf{P (Pattern)}: empirically verifiable signals or trends predicted by the hypothesis.
   
\end{itemize}
We refer to this structure as the CAMP theory of urban hypotheses. It forms the backbone of our hypothesis generation process. 

Building on this structure, the Ideation Agent (as shown in Figure~\ref{fig:3}) performs four types of scientifically meaningful transformations to create diverse, domain-consistent, and potentially innovative hypotheses: 
\begin{itemize}

\item \textbf{Recombination of existing hypotheses.} 
By decomposing existing hypotheses into CAMP elements (context, variables, mechanisms, patterns), the agent recombines elements across domains, generating new variable pairs, interdisciplinary connections, and novel conceptual pairings. This enables hypothesis creation beyond the limitations of human disciplinary silos.
\item \textbf{ Mechanism transformation.}
The agent explores alternative causal pathways by substituting known mechanisms with new mediators, moderators, or conditional processes. This is particularly important in urban systems, where multiple socioeconomic, environmental, and behavioral pathways often interact.
\item \textbf{Context transfer.}
The system can migrate hypotheses from one spatial or social context to another—from one city to another, from national to global scale, or across demographic or institutional environments. Contextual migration allows hypotheses to generalize or specialize, supporting comparative urban research.
 \item \textbf{Meta-hypothesis generation.}
Beyond generating direct hypotheses, the agent forms meta-hypotheses—higher-level scientific questions about mechanism validity, cross-regional differences, or system-level patterns. These meta-hypotheses guide new directions for exploration or deeper empirical inquiry.
\end{itemize}

Through these structured transformations, the Ideation Agent produces an initial idea set. These ideas are then passed into a debate stage involving multi-disciplinary Virtual Scientists, whose diverse perspectives help refine the conceptual plausibility. In parallel, the Critic Agent, fine-tuned on urban research review corpora, evaluates methodological soundness and theoretical rigor. Together, these iterative feedback loops support a dynamic hypothesis evolution process, converging on refined, domain-consistent hypotheses suitable for empirical testing.

\subsection*{Comprehensive Urban Data Search}


\begin{figure*}[t]
    \centering
    \includegraphics[width=0.9\linewidth]{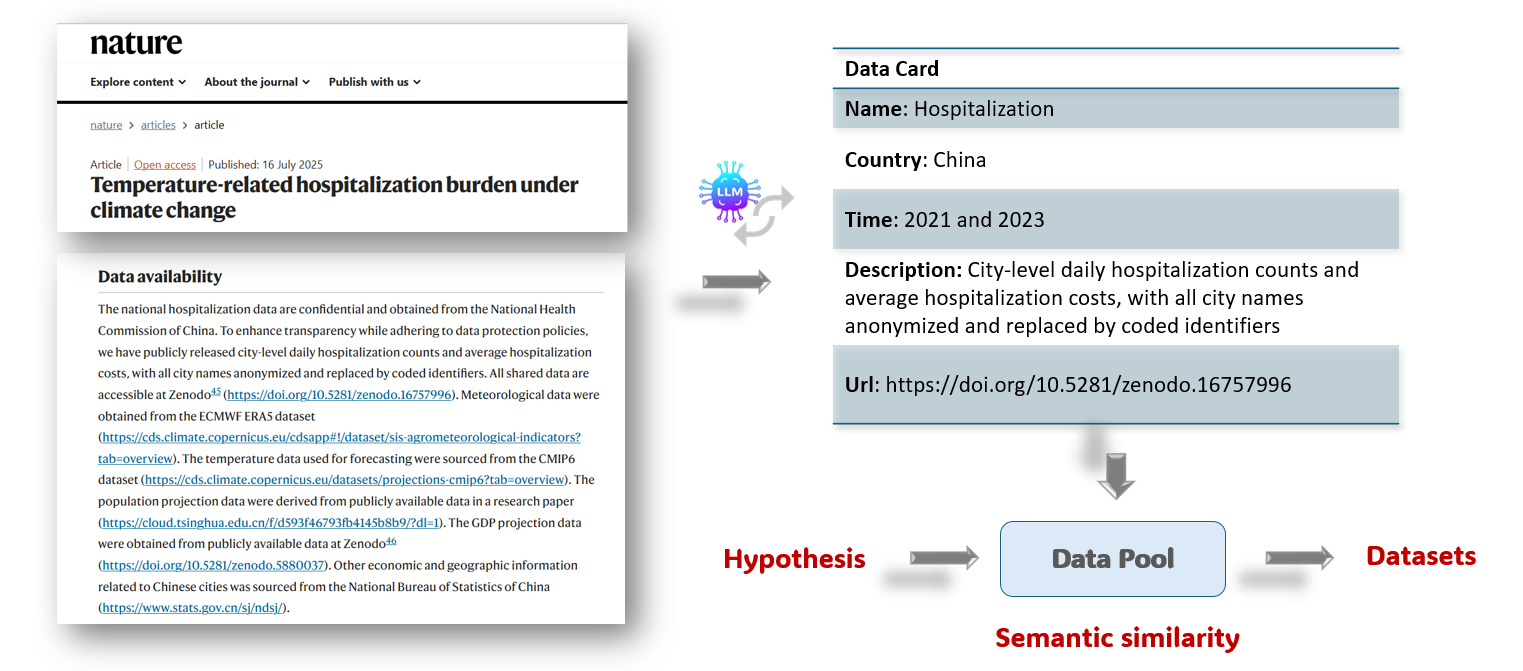}
    \caption{\textbf{The dataset construction and retrieval workflow of the Data Agent.}
The Data Agent extracts dataset information directly from the Data Availability sections of urban research papers using LLM-based semantic parsing, producing standardized data cards that include dataset name, region, time period, description, and URL. These cards populate a unified data pool spanning four major urban data categories: statistical infrastructure, human behavior, policy and survey, and multimodal sensing. Through semantic similarity matching, the Data Agent links hypotheses to relevant datasets and enables automated downloading, preprocessing, and integration for downstream analysis. }
    \label{fig:5}
\end{figure*}

High quality empirical urban research depends on diverse data sources that cover infrastructure, mobility, environment, economy, health, policy, and human behaviour. However, many high value datasets are not directly discoverable through standard web search. They are often only mentioned in the data availability sections of academic articles, embedded in supplementary materials, or hosted on domain specific repositories and institutional portals. To address this gap, our AI urban scientist includes a Data Agent (Figure~\ref{fig:5}) that is responsible for discovering, structuring, and retrieving urban datasets in a unified way.

\paragraph{Extracting datasets from literature and web sources.}
The Data Agent first scans scientific articles, with a focus on their data availability statements and related sections. Using the semantic retrieval and information extraction capabilities of large language models, it identifies text segments that describe datasets, including dataset names, spatial coverage, temporal range, measurement types, and access links. Based on more than 2{,}000 domain papers, the agent converts these unstructured descriptions into structured \emph{data cards}. 

\begin{table}[t]
\centering
\caption{Schema of structured data cards used to index urban datasets.}
\label{tab:data_card_schema}
\begin{tabular}{p{3.2cm} p{10.5cm}}
\toprule
\textbf{Field} & \textbf{Description} \\
\midrule
Name & Dataset name or short descriptive title. \\
Country / Region & Primary spatial coverage of the dataset. \\
Time & Observation period or reference years covered by the data. \\
Type & Data category and subcategory (e.g., environmental, socioeconomic, mobility). \\
Description & Brief summary of key variables, spatial and temporal granularity, and typical use cases. \\
URL & Link to the data repository, DOI landing page, or official access portal. \\
\bottomrule
\end{tabular}
\end{table}

As show in Table~\ref{tab:data_card_schema}, these data cards form a machine readable index that populates a unified data pool.

\paragraph{Integrating private and institutional data sources.}
Urban research frequently relies on non public or partially restricted data, such as hospital records, administrative microdata, or operator specific sensor networks. The Data Agent extends the above pipeline to internal and institutional databases by applying the same data card abstraction to local catalogues and documentation. This allows hypotheses to be linked not only to open datasets but also to private data sources, while respecting access control policies.

\paragraph{Automatic download, preprocessing, and integration.}
Once a relevant dataset has been identified, the Data Agent attempts to automate the main technical steps required for empirical analysis. Where possible, it
\begin{itemize}
    \item downloads data through APIs, DOIs, or direct links;
    \item performs basic preprocessing such as format conversion, cleaning, spatial joins, temporal alignment, and unit standardisation;
    \item integrates heterogeneous datasets into a common structure suitable for downstream modelling.
\end{itemize}
By automating these routine but time consuming tasks, the Data Agent substantially reduces the friction between hypothesis generation and empirical testing.

\paragraph{A taxonomy of urban data types.}
The datasets collected by the Data Agent span four major categories, which together cover the core components of urban systems:
\begin{enumerate}
    \item \textbf{Statistical infrastructure data}: road networks and transportation infrastructure; building footprints and land use maps; points of interest; administrative boundaries and zoning maps; utility networks for electricity, water, and communication.
    \item \textbf{Human behaviour data}: human mobility traces (GPS, transit card, ride hailing); socioeconomic activities such as consumption, employment, and commerce; social media interactions and online behaviour; health and wellbeing data, including hospitalisation counts and survey based indicators.
    \item \textbf{Policy and survey data}: population census and household surveys; statistical yearbooks and socioeconomic indicators; government reports and planning documents; policy texts and regulatory frameworks.
    \item \textbf{Multimodal sensing data}: satellite remote sensing imagery (optical, SAR, night lights); aerial and drone imagery; ground based sensors for air quality, temperature, and noise; urban IoT devices monitoring traffic, energy, and water; city wide camera networks and meteorological stations.
\end{enumerate}
This taxonomy supports structured indexing and allows the system to retrieve datasets that are compatible with a given hypothesis in terms of variables, mechanisms, and context.

\paragraph{Semantic similarity between hypotheses and datasets.}
Both hypotheses and data cards are embedded into a shared semantic space. For any given hypothesis, the Data Agent retrieves the most relevant datasets by computing semantic similarity between the CAMP representation of the hypothesis and the textual descriptions of candidate datasets. This matching takes into account variables, mechanisms, spatial and temporal context, and data modality. In this way, the Data Agent can recommend empirically suitable datasets and provide the data backbone needed for downstream analysis by the AI urban scientist.

\subsection*{Automatic Data Analysis}


\begin{figure*}[t]
    \centering
    \includegraphics[width=0.9\linewidth]{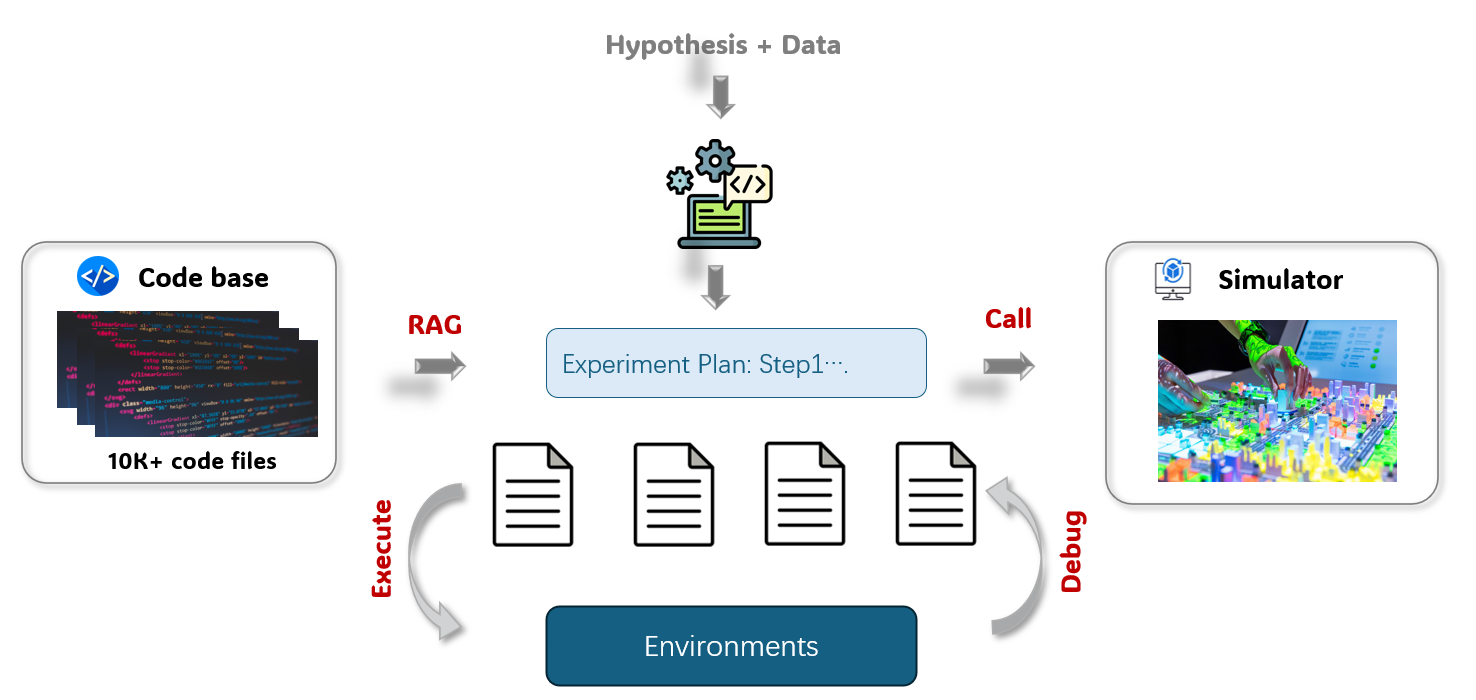}
    \caption{\textbf{Workflow of the Automatic Data Analysis Agent.} The agent receives a hypothesis–dataset pair and generates an experiment plan by retrieving relevant examples from a code base of 10K+ scripts. It then produces executable multi-language code, runs the scripts in controlled environments, and iteratively debugs errors. When empirical data are insufficient, the agent can call external simulators to generate synthetic evidence. Together, these capabilities enable end-to-end empirical testing for heterogeneous urban research tasks.}
    \label{fig:6}
\end{figure*}

urban research requires the integration of highly heterogeneous datasets and the execution of diverse analytical tasks, ranging from high-dimensional statistical modelling to machine learning, spatial analysis, simulation, and causal inference. These tasks often involve multiple programming languages (e.g., Python, R) and specialised software environments. However, existing data–science agents are far from capable of handling such complexity: they can neither design multi-step analysis pipelines nor reliably generate executable code across heterogeneous tools and data modalities. To address this gap, we introduce an \emph{Automatic Data Analysis Agent} (Figure~\ref{fig:6}) that performs end-to-end empirical analysis for a given hypothesis--dataset pair.

\paragraph{Knowledge-driven task planning.}
Rather than relying solely on prompting or ad-hoc generation, the agent is grounded in a code knowledge base containing more than 10{,}000 scripts extracted from published urban research papers. These scripts cover common analytical tasks such as regression models, spatiotemporal analysis, exposure–response modelling, machine learning prediction, uncertainty quantification, and robustness checks. By retrieving relevant examples from this code base through RAG (retrieval-augmented generation), the agent constructs an \emph{experiment plan} that outlines the analytical steps required to test the hypothesis.

\paragraph{Multi-language code generation and execution.}
Based on the retrieved templates and the structured description of the dataset, the agent generates executable code across Python, R, or other required languages. Each step of the experiment plan is translated into runnable scripts. The system then executes the generated code in isolated computational environments, capturing outputs, warnings, and errors.

\paragraph{Iterative error diagnosis and correction.}
Real-world urban datasets are noisy, inconsistent, and often incomplete. As a result, code execution naturally triggers errors. The agent includes an automatic debugging loop: execution logs are fed back into the LLM, which corrects syntax errors, resolves missing-library issues, adjusts data cleaning steps, and fixes incorrect assumptions about variable formats or dimensionality. This iterative loop continues until the pipeline executes successfully, producing empirical results.

\paragraph{Simulation-based hypothesis testing.}
In many urban research problems, empirical data may be missing, insufficient, or unsuitable for directly testing a hypothesis. To address such cases, the agent can call external simulators---such as climate–health models, mobility simulators, agent-based city models, or environmental exposure models---by generating the required configuration files and code wrappers. The simulator output is then integrated into the analysis pipeline, enabling counterfactual reasoning or synthetic experiments when observational data are limited.

\paragraph{Unified execution environment.}
Finally, all code is executed within controlled computational environments that include standard statistical libraries, geospatial toolkits, and machine learning frameworks. This ensures reproducibility and facilitates cross-language execution while shielding the user from the complexity of environment management.

Through these stages---knowledge-driven planning, retrieval-based code guidance, multi-language code generation, error correction, and simulator integration---the Automatic Data Analysis Agent provides a practical and robust mechanism for end-to-end empirical urban research research.

\subsection*{Reliable Urban Research Critic}

To ensure that the hypotheses generated by the Ideation Agent are evaluated through a lens consistent with the standards of high-quality urban research, we develop a Critic Agent that emulates the judgment patterns of expert reviewers in the field. Unlike general-purpose AI reviewers, which tend to inherit evaluation criteria from machine learning venues such as OpenReview, our Critic Agent is explicitly grounded in the epistemic standards of Nature and Nature Cities, two flagship journals representing the highest level of urban and interdisciplinary research.  As shown in Figure~\ref{fig:3}, the key steps of constructing the Critic Agent are as follows. 

\paragraph{System-Level Alignment with urban research Standards.}
We begin by constructing a domain-specific system prompt that integrates the editorial requirements, aims and scope, and mission statements of Nature and Nature Cities. These statements emphasize conceptual significance, interdisciplinary depth, empirical grounding, methodological rigor, and societal relevance. By embedding these principles directly into the system prompt, we ensure that the Critic Agent evaluates hypotheses according to norms that are aligned with the expectations of top-tier urban research journals.

\paragraph{Fine-Tuning with Expert Review Corpora.}
Next, we curate a dataset of 2,000+ publicly available peer-review reports from the Nature family journals. These reviews contain both positive and negative assessments across dimensions such as conceptual novelty, methodological transparency, evidence sufficiency, robustness, and policy relevance. We decompose reviewer comments into structured labels (e.g., strengths, weaknesses, major concerns, minor concerns) and use them to fine-tune the model. This process teaches the Critic Agent not only what constitutes strong scientific reasoning, but how expert reviewers articulate critiques, balance strengths and weaknesses, and justify their judgments.

\paragraph{Impact-Factor–Aligned Evaluation Calibration.}
To further calibrate the model’s evaluative behavior, we leverage the 15,000 hypotheses extracted from domain literature, assigning each hypothesis a quality tier (Tier 1–4) based on the impact factor of the venue in which the original paper was published. This allows us to augment the fine-tuning dataset with a large corpus of labeled examples that reflect real-world hierarchical standards of scientific merit. By incorporating this multi-level calibration, the Critic Agent learns to distinguish between high-impact, field-shaping ideas and lower-tier, incremental hypotheses.

Through system prompt alignment, expert-review fine-tuning, and impact-factor–based calibration, the Critic Agent develops evaluation capabilities that reflect the values of the urban research community. It can judge whether a hypothesis has conceptual significance, cross-disciplinary relevance, empirical feasibility, and theoretical credibility. Its output includes both a tier rating (Tier 1–4 or reject) and explanatory feedback, enabling the hypothesis refinement loop to converge toward scientifically meaningful ideas.

\begin{figure*}[t]
    \centering
    \includegraphics[width=0.8\linewidth]{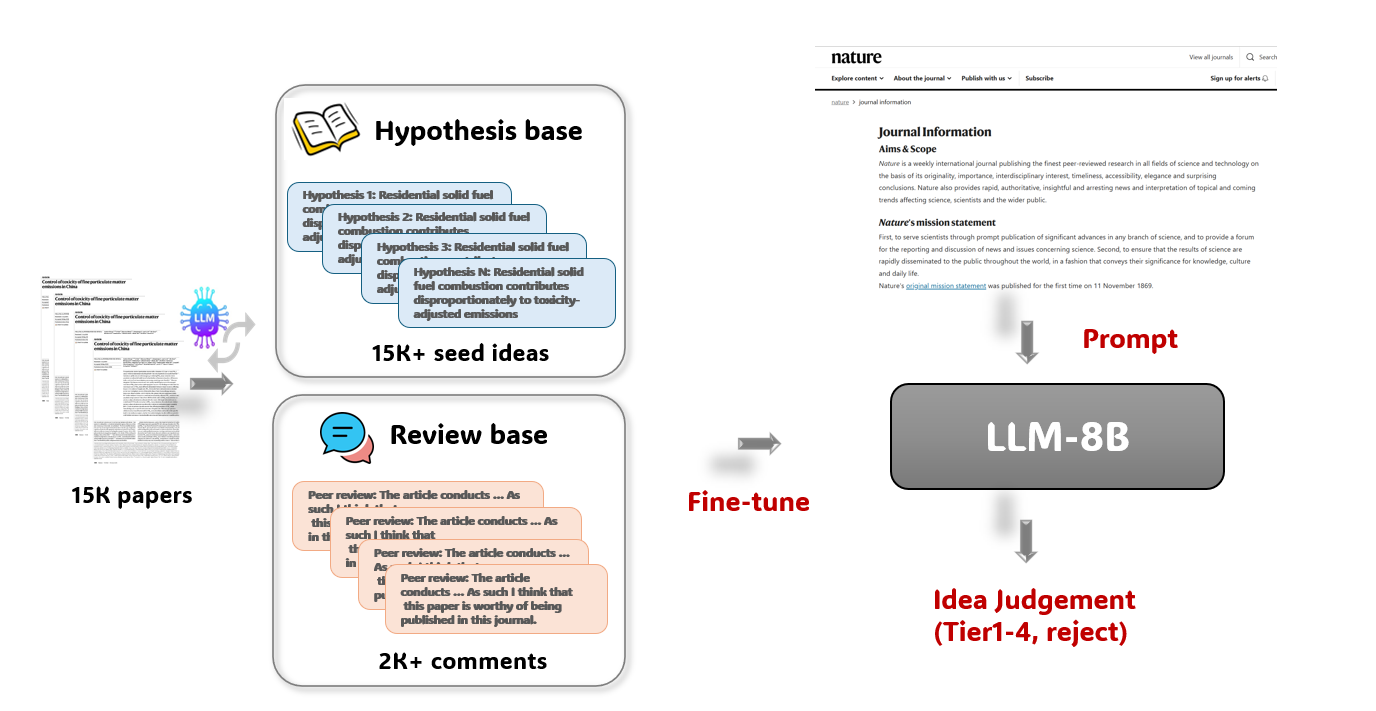}
    \caption{\textbf{The construction and training pipeline of the Critic Agent.}
We align the Critic Agent with domain standards by integrating the editorial principles of Nature and Nature Cities as the system prompt. A corpus of 15K papers provides 15K+ seed hypotheses, while 2K+ expert reviewer comments from Nature family journals supply high-quality positive and negative evaluation signals for fine-tuning. Additionally, hypotheses extracted from the 15K papers are labeled into four tiers according to journal impact factor, enabling multi-level calibration. Together, these components produce a domain-informed Critic Agent capable of assigning idea-quality tiers (Tier 1–4 or reject) consistent with expert judgment in urban research. }
    \label{fig:4}
\end{figure*}



\section*{Preliminary  Results}

Next, we present preliminary experimental results from the individual agents that constitute our AI Urban Scientist system.

\subsection*{Examples of Generated Hypotheses}

To illustrate how the Ideation Agent synthesizes novel hypotheses through structured recombination, we present below an example generated from the CAMP framework. The agent retrieves relevant concepts across financial inclusion, spatial inequality and urban economic dynamics, identifies complementary mechanisms, and recombines them into a coherent, testable hypothesis. Notably, a literature search conducted by the system indicates that no existing urban science study has examined the dual effect of mobile money on spatial inequality—initial decentralization followed by renewed polarization—suggesting that the generated idea occupies a genuinely unexplored space in current research.

\begin{lstlisting}
{
  "New CAMP": {
    "Context": "Emerging economies with rapid urbanization and digital financial inclusion (India, Kenya, Nigeria), 2010-2025.",
    "A": "Mobile money penetration (active accounts per capita, transaction volumes).",
    "B": "Urban spatial inequality (dispersion of nighttime lights, sub-city GDP).",
    "Mechanism": "Mobile money lowers transaction costs and decentralizes activity, but also strengthens agglomeration via faster capital circulation in core zones.",
    "Pattern": "Initial decline in inequality (2010-2020), followed by polarization as digital liquidity concentrates in selected hubs.",
    "InnovationType": "Recombination",
    "Innovation rationale":  "Combines financial integration (from Hypothesis 2) and institutional quality (from Hypothesis 3) but applies them to urban sustainability (a novel B variable).",
    "WhyItMatters": "Reveals dual effects of digital finance-decentralization and concentrated liquidity-informing inclusive urban-digital policy."
  },
  
  "New Idea": {
    "Title": "Mobile Money and Urban Spatial Inequality: A Dual-Pathway Analysis (2010-2025)",
    "Abstract": "We study how mobile money penetration shapes spatial inequality in emerging economies. Combining digital finance indicators and geospatial measures (nighttime lights, sub-city GDP), we hypothesize a dual mechanism: decentralization through lower participation barriers, and strengthened agglomeration via accelerated capital circulation. Using data from India, Kenya and Nigeria, we test for a temporal shift from reduced inequality (2010-2020) to renewed polarization post-2020. Results inform debates on digital finance's redistributive effects and inclusive urban development."
  }
}
\end{lstlisting}

\subsection*{Examples of Hypotheses Review}

While recent advances in large language model-based reviewer systems such as DeepReviewer-14B (ACL 2025) demonstrate impressive capabilities in standardizing and accelerating peer-review workflows, their performance can vary across scientific domains. As an illustrative example, we examined a Nature research article on temperature-related hospitalization risks. DeepReviewer-14B, trained primarily on machine-learning manuscripts and OpenReview-style evaluations, classified the study as a Reject. A closer inspection suggests that this outcome likely reflects differences in disciplinary norms rather than deficiencies in the paper itself: Urban Science and climate-health research often emphasize context-specific evidence, observational data quality, spatial heterogeneity, and policy relevance—dimensions that are less central in ML-oriented evaluation frameworks focusing on algorithmic novelty, generalization benchmarks, and ablation-based validation.
\newpage

\begin{lstlisting}
{
  "Paper": {
    "Title": "Temperature-related hospitalization burden under climate change",
    "Journal": "Nature",
    "PublishTime": "2025-07-16",
    "Topic": "Extreme temperature, hospitalization, climate scenarios, China"
  },

  "Review": {
    "Decision": "Reject",
    "Rating": 6.0,
    "Soundness": 2.75,
    "Presentation": 3.0,
    "Contribution": 2.75
  },

  "Summary": {
    "MainFinding": "Estimates current and future temperature-related hospitalization risks in 301 Chinese cities (2021-2023), under SSP1-2.6, SSP2-4.5, SSP5-8.5.",
    "Methods": "DLNM for temperature-hospitalization relationships; projections to 2100 with climate scenarios.",
    "EconomicIndex": "Introduces Hospitalization Burden Economic Index (HBEI) to quantify economic burden of temperature-related hospitalizations."
  },

  "Strengths": [
    "Large, city-level dataset from >7,000 hospitals across 301 cities (~90% of Chinese cities).",
    "Forward-looking projections of hospitalization risks and burdens to 2100 under multiple SSP scenarios.",
    "Novel HBEI index that links health impacts with economic burden and urban development context.",
    "Detailed spatial patterns and age-specific vulnerabilities (children, adolescents, elderly).",
    "Clear policy implications for city-level assessments and targeted adaptation strategies."
  ],

  "Weaknesses": [
    "Very short observation window (2021-2023) limits ability to capture long-term trends and past extreme events.",
    "Adaptation is only partially modeled via dynamic thresholds; does not explicitly include behavioural or infrastructural adaptation.",
    "Focuses solely on hospitalizations, omitting mortality, outpatient visits and broader health outcomes.",
    "Results may have limited generalizability beyond China due to context-specific demographic and healthcare features.",
    "Reliance on daily mean temperature may oversimplify heat-health relationships; alternative metrics (heat index, wet-bulb) not explored."
  ],

  "Suggestions": [
    "Extend the historical period (multi-decade data) and incorporate lagged effects of temperature on health.",
    "Model explicit adaptation measures (AC use, urban design, early-warning systems) and potential acclimatization.",
    "Include additional health outcomes (mortality, morbidity, outpatient visits) and indirect effects (air quality, vector-borne diseases).",
    "Replicate analyses in other regions and include confounders such as air pollution for broader generalizability.",
    "Use more nuanced thermal metrics (heat index, wet-bulb) and assess uncertainty across multiple climate models."
  ],

  % "OpenQuestions": [
  %   "How would conclusions change with a longer time series and evolving vulnerability/adaptation?",
  %   "Does the current adaptation proxy (dynamic T2 threshold) adequately capture acclimatization?",
  %   "How robust are projections to uncertainties in climate models and scenarios?",
  %   "Why was daily mean temperature chosen over alternative thermal metrics, and how sensitive are findings to this choice?", "What are the unique advantages and limitations of HBEI compared with existing economic burden indices?"
  % ]
}
\end{lstlisting}

By contrast, our Urban Critic Agent, which is trained on domain-aligned review corpora from Nature, Nature Cities, and climate-health literature, evaluated the same study as a Tier-1, high-impact contribution. The agent highlighted the significance of national-scale health data, long-horizon climate projections, the development of a new economic burden index, and the policy importance of regional vulnerability patterns. This comparison illustrates that effective assessment of Urban Science research requires epistemic criteria tailored to its unique data structures, methodological traditions, and real-world objectives. Rather than emphasizing incompatibility, this case underscores the importance of domain-specific reviewer models that complement general AI reviewers and ensure fair, context-aware evaluation across diverse scientific fields.

\subsection*{Examples of Extracted Datasets}

The following panel summarizes the datasets extracted by our system. These resources can be queried via structured labels or natural-language search, providing streamlined and versatile access for a wide range of analytical tasks.

\begin{figure*}[htbp]
    \centering
    \includegraphics[width=0.99\linewidth]{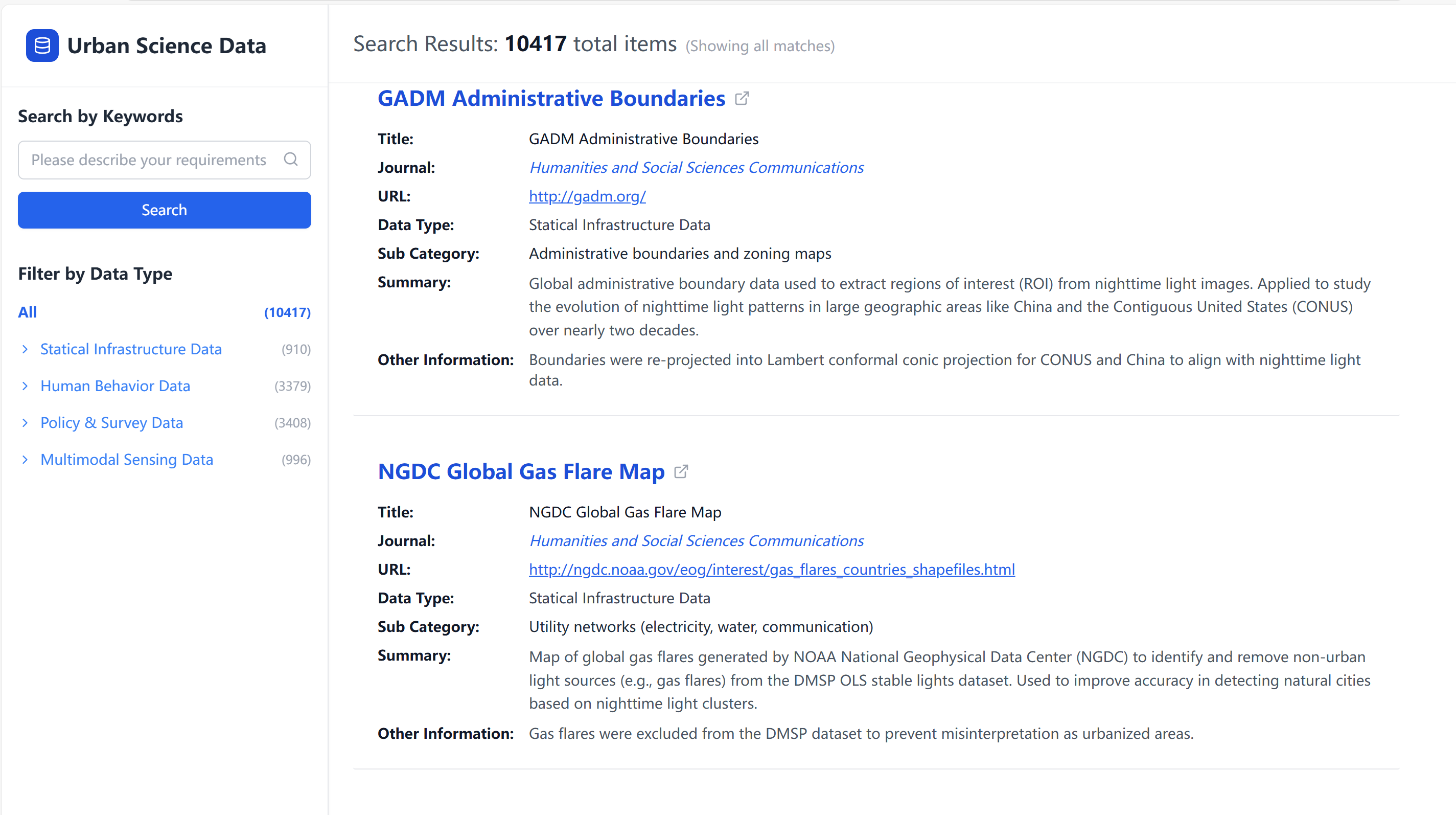}
    \caption{\textbf{Overview of the extracted datasets.} Summary of extracted datasets, accessible through both label search and natural-language queries. 
    } 
    \label{fig:7}
\end{figure*}

\subsection*{Examples of Experiment Conduction}

To evaluate the capability of the Analyzing Agent, we applied it to the paper “Using machine learning to assess the livelihood impact of electricity access”~\cite{Ratledge2022}. 

The agent automatically decomposed the entire study into a seven-stage, fully modularized software blueprint that spans the complete research workflow—from creating an empty project directory to reproducing the paper’s causal estimates with comprehensive testing and reproducibility checks. Across these seven phases, the agent specifies the full engineering pipeline: project initialization and configuration; multi-source data preprocessing and harmonization; satellite-image processing and CNN model development; wealth prediction and causal inference (DiD, matrix completion, synthetic control); confounding and mechanism analysis; visualization and figure generation; and, finally, systematic unit tests, integration tests, replication of published results, and reproducibility validation through configuration sweeps.

Overall, the output demonstrates that the Analyzing Agent can translate a complex ML-and-causal-inference study into an actionable, end-to-end software design, with clearly defined modules, inputs, outputs, and validation criteria. Rather than producing only code snippets, the agent constructs a coherent engineering plan that mirrors best practices in scientific software development, enabling faithful reproduction, systematic experimentation, and transparent extension of existing urban science research.

\begin{lstlisting}
{
 "Paper": "Using machine learning to assess the livelihood impact of electricity access",
  "AnalyzingAgentPlan": {
    "Phase1_ProjectSetup": {
      "tasks": [
        "Create project structure (src/, data/, configs/, tests/)",
        "Install Python/R dependencies",
        "Implement YAML-based config manager"
      ]
    },

    "Phase2_DataProcessing": {
      "tasks": [
        "Compute PCA-based asset wealth index",
        "Digitize and harmonize electricity grid maps",
        "Identify treatment/control via spatial buffers",
        "Construct modeling-ready integrated datasets"
      ]
    },

    "Phase3_SatelliteAndCNN": {
      "tasks": [
        "Generate multispectral composites with cloud masks",
        "Modify ResNet-18 for 6-channel regression",
        "Implement custom loss (MSE + bias penalty)",
        "Build CNN training pipeline with metrics"
      ]
    },

    "Phase4_WealthImputation_CausalInference": {
      "tasks": [
        "Predict wealth panel using trained CNN",
        "Implement DiD, Matrix-Completion, Synthetic Control",
        "Run bootstrap uncertainty estimation"
      ]
    },

    "Phase5_Confounding_AssetImpact": {
      "tasks": [
        "Analyze infrastructure covariates (roads, cell networks)",
        "Conduct asset-level DiD impact analysis"
      ]
    },

    "Phase6_Visualization": {
      "tasks": [
        "Generate maps, effect curves, validation plots"
      ]
    },

    "Phase7_Testing_Reproducibility": {
      "tasks": [
        "Unit tests and integration tests",
        "Validate outputs against paper results",
        "Reproducibility checks via config variations"
      ]
    }
  }
}
\end{lstlisting}

\section*{Discussion and Future Work}

\begin{figure*}[t]
    \centering
    \includegraphics[width=0.99\linewidth]{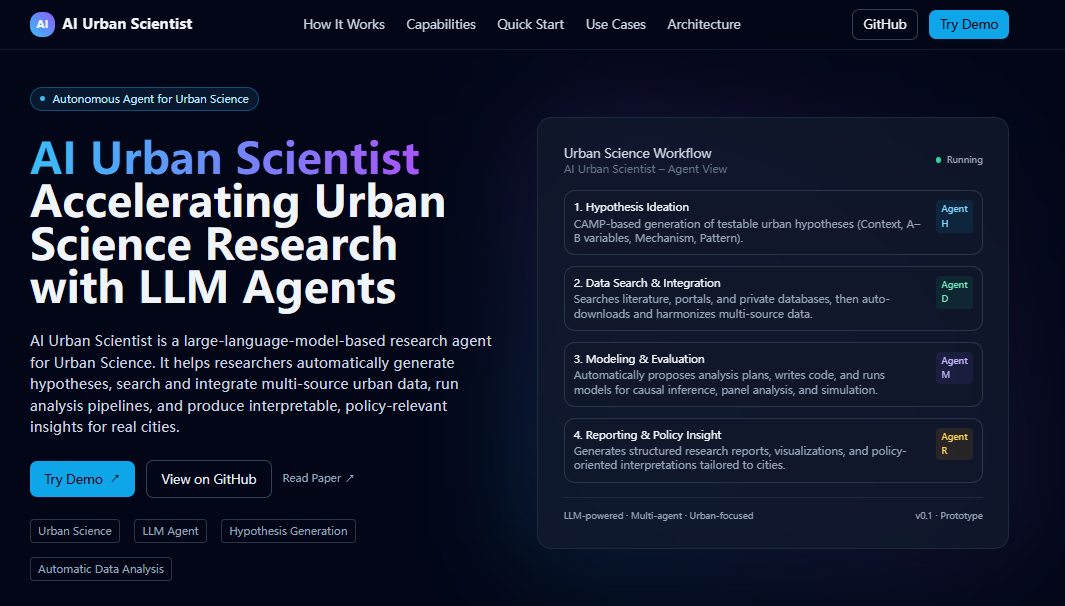}
    \caption{\textbf{Overview of the AI Urban Scientist platform.} The system integrates four core agents-Hypothesis Ideation, Data Search and Integration, Modeling and Evaluation, and Reporting and Policy Insight-into a unified, easy-to-use interface. 
    The platform is designed not only to make these tool agents accessible to researchers, but also to support community contributions and the sharing of new tools, datasets, and analytical components. This enables AI Urban Scientist to evolve into an open, extensible ecosystem for automated urban science research.
    } 
    \label{fig:8}
\end{figure*}

Despite the promising capabilities demonstrated by AI Urban Scientist, several key challenges remain before such systems can be widely and reliably adopted across the urban science community. Addressing these challenges requires not only technical innovation but also collaborative participation from researchers, institutions, and policy analysts.

\subsection*{Building systematic evaluation benchmarks}

A major limitation is the absence of quantitative benchmarks for evaluating each stage of the research workflow. Urban science does not yet have standardized tests for hypothesis quality, dataset relevance, modeling correctness, or policy interpretation. This makes it difficult to compare system designs, measure progress, or build confidence in automated research agents.

To move forward, the field needs open and transparent evaluation resources-such as hypothesis-data matching benchmarks, reproducible analysis tasks, and reviewer-style rubrics aligned with journals like Nature Cities. As part of this effort, the tools provided on our platform offer a starting point for experimenting with structured tasks and standardized workflows. We encourage researchers to use the website, test the agents on their own problems, and contribute feedback that can guide the development of formal benchmarks.

\subsection*{Human roles in AI-driven research: from co-pilot to supervisory oversight}

Another central question concerns the evolving role of human researchers. At present, AI Urban Scientist functions as a co-pilot: it assists with ideation, dataset search, analysis planning, and code execution. As capabilities grow, human involvement may shift toward a supervisory role-verifying outputs, examining assumptions, and providing high-level scientific judgement.

Issues of authorship, accountability, and interpretability will become increasingly important, especially in urban science, where conclusions influence real-world planning and policy. By using our system in real workflows, researchers can help define best practices and clarify where AI assistance should augment, rather than replace, human expertise. The platform is designed to support this kind of reflective experimentation, enabling users to explore how humans and agents can collaborate responsibly.

\subsection*{Toward standards and community-driven infrastructure}

The long-term success of AI-driven research in urban science depends on shared standards and community-maintained resources. Although our prototype integrates structured hypothesis representations (CAMP), dataset cards, code bases, and simulations, the field still lacks unified conventions for annotating datasets, describing mechanisms, or evaluating analytical pipelines.

We believe that progress requires a collaborative, open ecosystem-similar to how tool hubs and model repositories accelerated machine learning research. To this end, our website (Figure~\ref{fig:8}) provides an open interface where researchers can explore the system, run agents on real urban problems, and contribute datasets, code templates, or domain critiques. These community contributions will be essential for improving coverage, ensuring transparency, and enabling reproducible workflows.

We invite the urban science community to engage with the platform, test its capabilities, and participate in building shared knowledge bases and standards. Such collective efforts will help transform AI Urban Scientist from a single prototype into a broadly useful research infrastructure for automated, scalable, and interpretable urban science discovery.

\bibliography{reference}

\end{document}